\documentclass[aps,prl,twocolumn,showpacs,floatfix]{revtex4}
\usepackage{graphicx}
\begin{document}

\title{Entanglement in spatially inhomogeneous many-fermion systems}

\author{V. V. Fran\c{c}a and K. Capelle}
\affiliation{Departamento de F\'{\i}sica e Inform\'{a}tica,
Instituto de F\'{\i}sica de S\~ao Carlos, Universidade de S\~ao Paulo,
Caixa Postal 369, 13560-970 S\~ao Carlos, SP, Brazil}
\date{\today}

\begin{abstract}
We investigate entanglement of strongly interacting fermions in spatially
inhomogeneous environments. To quantify entanglement in the presence of 
spatial inhomogeneity, we propose a local-density approximation (LDA) to 
the entanglement entropy, and a {\em nested LDA} scheme to evaluate the 
entanglement entropy on inhomogeneous density profiles. These ideas are
applied to models of electrons in superlattice structures with different 
modulation patterns, electrons in a metallic wire in the presence of 
impurities, and phase-separated states in harmonically confined many-fermion 
systems, such as electrons in quantum dots and atoms in optical traps. We 
find that the entanglement entropy of inhomogeneous systems is strikingly 
different from that of homogeneous systems. 
\end{abstract}


\maketitle

\newcommand{\be}{\begin{equation}}
\newcommand{\ee}{\end{equation}}
\newcommand{\bea}{\begin{eqnarray}}
\newcommand{\eea}{\end{eqnarray}}
\newcommand{\bi}{\bibitem}
\newcommand{\la}{\langle}
\newcommand{\ra}{\rangle}
\newcommand{\ua}{\uparrow}
\newcommand{\da}{\downarrow}
\renewcommand{\r}{({\bf r})}
\newcommand{\rp}{({\bf r'})}
\newcommand{\rpp}{({\bf r''})}

An intense interdisciplinary research effort is currently directed at analyzing
entanglement in a wide variety of physical systems. Apart from the intrinsic 
interest in entanglement as one of the most counterintuitive predictions of 
quantum mechanics, this research effort is largely motivated by the prospect 
of simulating or exploiting entanglement as a resource for quantum information 
processing and computing \cite{nielsen}.

Particularly promising systems for quantum computation are solid-state
devices, because of their scalability and the possible integration with
existing silicon-based technology. However, real solids and solid-state 
devices are necessarily 
inhomogeneous, {\em i.e.}, characterized by boundaries, interfaces, spatial 
modulations of system parameters, impurities, defects, and externally applied 
fields (arising, {\em e.g.}, from gate electrodes or from spatially confining 
potentials), among others. Systems of optically trapped atoms are also 
necessarily inhomogeneous, due to the trapping potential. If quantum 
information processing or  computing is ever to become reality outside
the laboratory, we must be able to quantify entanglement in realistic, 
spatially inhomogeneous situations. 
Unfortunately, a simple and reliable prescription to quantify and interpret 
the degree of entanglement in inhomogeneous systems is still missing. Here 
we propose a solution to this problem in the context of the entanglement 
entropy.

The entanglement entropy is defined for a system divided in two subsystems, $A$
and $B$, as ${\cal E}=Tr_A (\rho_A \log_2 \rho_A)$, where $\rho_A=Tr_B \rho$ is
the reduced density matrix of subsystem $A$, and $\rho$ is the density matrix 
of the full system. The Hohenberg-Kohn theorem \cite{hk,kohnrmp,dftbook} 
guarantees that ${\cal E}$ is a functional of the ground-state 
density, ${\cal E}[n(x)]$, where $x$ represents sites or spatial coordinates 
and $n(x)$ the spatial distribution of particles over these sites, but that 
functional is not known in general. In the following, we make 
the realistic assumption that $\rho$ or ${\cal E}$ can be calculated, at least 
approximately, for a spatially uniform system, such as a lattice in which all 
sites are equivalent, or a uniform continuum system without boundaries. In
such spatially homogeneous systems, the density $n$ reduces to a constant,
and the functional ${\cal E}[n(x)]$ becomes the function ${\cal E}^{hom}(n)$.
Our aim is to develop a simple, yet reliable, approximation scheme for 
obtaining the ground-state entanglement entropy of inhomogeneous systems, 
${\cal E}[n(x)]$.

To this end we propose the local-density approximation
\be
{\cal E}[n(x)] \approx {\cal E}^{LDA}[n(x)] 
= \int dx\, {\cal E}^{hom}(n)_{n\to n(x)}, 
\label{slda}
\ee
in which the functional ${\cal E}[n(x)]$ is approximated by evaluating the 
per-site entropy of the homogeneous system, ${\cal E}^{hom}(n)$ at the 
density distribution of the inhomogeneous system. This local-density 
approximation (LDA) for the 
entanglement entropy is inspired by the LDA made in density-functional theory 
(DFT) for the exchange-correlation energy \cite{kohnrmp,dftbook}. 
There is also a conceptual relation to the LDA for the thermodynamic entropy 
made in classical DFT of inhomogeneous liquids \cite{tdentr} and 
in DFT for thermal ensembles \cite{ogk}.

\begin{figure*}
\begin{center}
\includegraphics[height=80mm,width=180mm,angle=0]{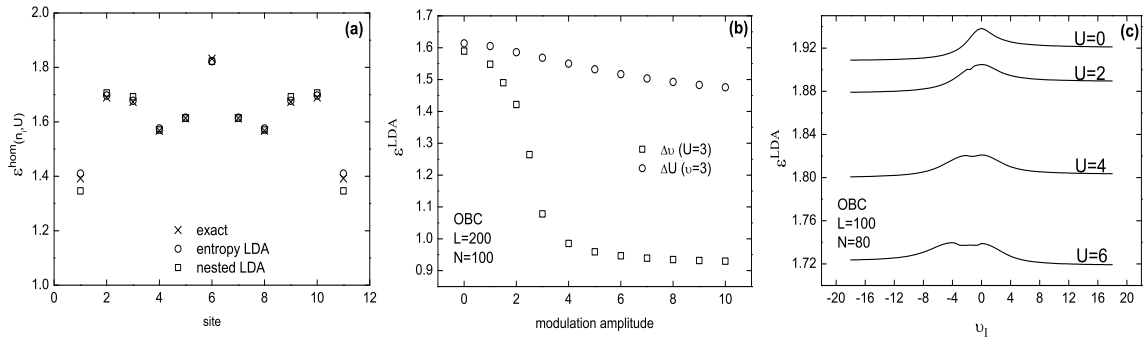}
\end{center}
\caption{\label{fig1} Panel (a):
Site-resolved entanglement entropy ${\cal E}^{hom}(n_i,U)
={\cal E}^{hom}(n,U)|_{n\to n_i}$ of an open 
Hubbard chain with impurity potential $v_I=-1$ on the central site. Crosses:
exact data for ${\cal E}(n_i,U)$, obtained by fully numerical diagonalization 
of the Hamiltonian.  
Circles: ${\cal E}^{hom}(n_i,U)$ evaluated on exact densities.
Squares: ${\cal E}^{hom}(n_i,U)$ evaluated on BA-LDA densities.
Panel (b): Average per-site entropy ${\cal E}^{LDA}[n_i,U]/L$  
of two superlattices, evaluated on BA-LDA densities.
Circles: constant repulsive on-site potential $v=3$ and repulsive on-site
interaction modulated with amplitude $\Delta$. Squares: constant
repulsive on-site interaction $U=3$ and on-site potential modulated
with amplitude $\Delta$. In both cases the system has $L=200$ sites,
$N=100$ particles, and the periodicity of the modulation is $\Delta L=20$
lattice sites.
Panel (c):  Average per-site entropy ${\cal E}^{LDA}[n_i,U]/L$ 
of itinerant interacting particles
in the presence of a localized impurity of strength $v_I$. All
calculations were done for open boundary conditions (OBC), but results for
periodic boundary conditions are qualitatively the same.}
\end{figure*}

Clearly, the LDA for the entanglement entropy is a general concept, applicable
to a wide variety of systems, but to be specific, and to compare to previous 
work, we here consider interacting fermions on a lattice, described by 
the Hubbard model,
\be
\hat{H}= -t\sum_{i,\sigma} (c_{i\sigma}^\dagger c_{i+1,\sigma}+H.c.)
+U\sum_i \hat{n}_{i\uparrow} \hat{n}_{i\downarrow}
+ \sum_{i\sigma} v_{i\sigma}\hat{n}_{i\sigma}
\label{humo}
\ee
where $\hat{n}_{i\sigma}=c_{i\sigma}^\dagger c_{i\sigma}$, with 
$\sigma=\uparrow,\downarrow$, is the local density operator expressed in terms 
of fermionic creation and annihilation 
operators, $U$ is the on-site interaction, $t$ the inter-site hopping (below 
taken to be the unit of energy), and $v_{i\sigma}$ represents external electric
and magnetic fields. Inclusion of this last term makes the model inhomogeneous.
The HK theorem guarantees that the global density distribution $n={n_1,..,n_L}$ 
determines the ground-state wave function of the inhomogeneous $L$-site model.

The homogeneous Hubbard model ($v_{i\sigma}\equiv0$) accounts, approximately, 
for spin and charge degrees of freedom, for the itineracy of the charge 
carriers, and for their interactions. Its rich phase digram and nontrivial
physics, together with the available exact solution in one dimension, has
attracted much interest also in the quantum information community. In 
particular, several groups investigated entanglement measures for the Hubbard 
model (see, {\em e.g.}, Refs.~\cite{gu,johannesson,sarandy,korepin,zanardi,anfossi,vvfkc} and references therein). As a result, the single-site entanglement 
entropy (in which subsystem $A$ is one site) of the homogeneous Hubbard model 
in one dimension is known analytically \cite{gu,johannesson,vvfkc}:
\bea
{\cal E}^{hom}\left(n,U\right)=
-2\left({n\over 2}-\frac{\partial e}{\partial U}\right)
\log _{2}\left[{n\over 2}-\frac{\partial e}{\partial U}\right]\nonumber \\
-\left(1-n+\frac{\partial e}{\partial U}\right)
\log _{2}\left[1-n+\frac{\partial e}{\partial U}\right]
-\frac{\partial e}{\partial U}\log _{2}\left[\frac{\partial e}{\partial U}\right].
\label{tangle}
\eea
Here ${\cal E}^{hom}(n,U)$ is the single-site entanglement entropy 
(Shannon-von\,Neumann entropy of four probabilities) per site, and 
$e=E(n,U)/L$ is the 
ground-state energy per site, both as functions of interaction strength $U$ 
and particle density $n=N/L$, where $N$ and $L$ are the total 
number of particles and lattice sites, respectively. Since $e(n,U)$ 
can be obtained \cite{balda,trapprb} from the Bethe-Ansatz integral 
equations, Eq.~(\ref{tangle}) can
be used to extract detailed quantitative information about entanglement in
various phases and at the transitions between them \cite{gu,johannesson,vvfkc}.
However, this expression is valid only for the homogeneous Hubbard model,
{\em i.e.}, one in which all sites are equivalent. Such idealization is
most useful for analytical and numerical work, but provides only a rather 
imperfect picture of the physical situation in real solids or devices. 

To make use of expression (\ref{tangle}) for the entanglement entropy of
homogeneous Hubbard models also in inhomogeneous situations, we apply the 
local-density approximation (\ref{slda}) in the form
\be
{\cal E}[n_i,U] \approx {\cal E}^{LDA}[n_i,U] 
= \sum_i {\cal E}^{hom}(n,U)|_{n\to n_i}. 
\label{humolda}
\ee

For a given distribution $n_i$ of particles over sites,
Eq.~(\ref{humolda}) can be evaluated immediately. If, on the other hand, this
distribution is not known, it must be obtained in a separate calculation.
This is the case, e.g., if the system is specified by giving the potential
the particles move in, instead of the spatial distribution of particles. 
Experimentally, specifying the potential is more realistic, and we thus focus
on this case.

To obtain the ground-state charge and spin densities for a given external
potential, we again appeal to density-functional theory, this time employing 
the LDA in its standard formulation \cite{kohnrmp,dftbook}, {\em i.e.}, as an 
approximation to the exchange-correlation energy that is to be added to the 
mean-field energy functional. Minimization with respect to $n(x)$ then yields 
the ground-state energy and ground-state density profile. Specifically for the 
Hubbard model, we use the Bethe-Ansatz LDA (BA-LDA) of 
Refs.~\cite{balda,trapprb}. 
The combined calculation is thus a {\em nested LDA}, where BA-LDA is used to 
generate selfconsistent density profiles, and the entropy LDA (\ref{slda}) 
is used to predict the entropy on the resulting inhomogeneous density 
distribution. 

Both BA-LDA and the entropy LDA by construction become exact for a spatially
uniform system. BA-LDA has been shown by comparison to Monte Carlo, Bethe 
Ansatz, and density-matrix renormalization group data to provide density 
profiles that agree within a few percent with those of other many-body methods,
even for systems far from the uniform limit \cite{balda,trapprb,superl}. To 
quantify the reliability of the entropy LDA, we turn to small finite chains, 
which are far from the uniform limit, but for which the exact density matrix 
$\rho$ can be obtained by full numerical diagonalization of the Hamiltonian, so
that the exact entanglement entropy ${\cal E}$ is known.
Since small systems are a worst-case scenario for LDAs (which derive
from the thermodynamic limit), this is a particularly severe test for the
entropy LDA and nested LDA concepts.

Panel (a) of Fig.~\ref{fig1} displays the site-resolved entanglement 
entropy of an open chain chain with an impurity at the central site. 
Exact data are obtained by numerically diagonalizing the many-body Hamiltonian 
(\ref{humo}) and constructing the density matrix and the entanglement entropy 
from the exact eigenfunctions. This process also yields the exact density 
profile, which we use to evaluate the function ${\cal E}^{hom}\left(n,U\right)$
entering the entropy LDA, Eq.~(\ref{humolda}). The deviation between both sets 
of data measures the quality of the entropy LDA. In a separate calculation, we 
also generate BA-LDA density profiles \cite{balda,trapprb,superl}, and evaluate
 $ {\cal E}^{hom}\left(n,U\right)$ on them. This results in the nested LDA 
procedure, which can also be applied where the exact solution is not available.
As the figure shows, all three approaches agree closely, too within $1\%$, 
even far away from the limit in which LDAs become exact.

Having established the viability and reliability of the LDA for the 
entanglement entropy and of the nested LDA, we now turn to further 
experimentally relevant inhomogeneous systems. 
First, we investigate the effect a spatial modulation 
of system properties, in the form of a superlattice, has on the entanglement.
Naturally occuring or man-made systems displaying spatial modulations of
their properties on a nanoscale are important both as paradigms of simple
nanotechnological devices, and as particular examples of emerging spatial
inhomogeneity in strongly interacting many-electron systems. To model 
superlattice structures in the Hubbard model we follow Ref.~\cite{superl}, 
which, however, did consider only energies, not entropies.

Panel (b) of Fig.~\ref{fig1} displays the behaviour of ${\cal E}^{LDA}[n_i]$ 
for two
representative superlattice structures. According to the nested LDA concept,
the data were obtained by first running an ordinary BA-LDA calculation for a 
fixed distribution of on-site interaction and potentials, in order to generate 
the densities $n_i$. In a second step the resulting self-consistent density 
profile is substituted in Eq.~(\ref{slda}), in order to obtain 
${\cal E}^{LDA}[n_i]$, for the specified values of $U_i$ and $v_i$.
                                                                        
Two things leap to the eye in Fig.~\ref{fig1}(b): First, the larger the 
amplitude
of the modulation, the less entangled the system becomes. Indeed, the spatial 
inhomogeneity of the superlattice is expected to disrupt the entanglement 
present between itinerant charge carriers in the homogeneous system. 
Second, modulations in the on-site potential have a much more drastic effect 
on the entanglement entropy than those of the on-site interaction. 
Clearly, local fluctuations in the potential disentangle the system much
more efficiently than local fluctuations in the interaction. This is 
consistent with what was previously \cite{superl} observed for the 
ground-state energy, and again points to the importance of local electric 
fields in strongly correlated systems \cite{footnote}. It also means that
local electric fields are rather effective in reducing the degree of
entanglement between itinerant charge carriers.

Next, we turn to impurity systems. Specifically, we inquire what effect 
a local impurity potential has on the entanglement between itinerant
electrons. We model the impurity as $v_i = v_I \delta_{i,[L/2]}$, {\em i.e.,}
place a local electric potential on the central site of the L-site chain. 
Panel (c) of Fig.~\ref{fig1} shows that both for attractive ($v_I<0$) and 
repulsive ($v_I>0$) impurities, entanglement between the itinerant particles
is diminished, but by a much smaller margin than in the superlattice case,
where the potential is periodically repeated instead of acting on just 
one site. Stronger on-site interactions $U$ tend to freeze the density
distribution and thus reduce the number of degrees of freedom, which lowers
the entanglement entropy, as discussed for homogeneous systems in 
Ref.~\cite{vvfkc}. For small $U$, attractive impurities are more efficient
in disrupting itinerant entanglement than repulsive ones, but this difference
disappears for larger $U$. For attractive impurities, we observe a flat 
structure at small negative values of $v_I$. For these values, the average 
density of the impurity site remains fixed at $n_I=1$, which means that the
strenght of the impurity potential (which couples to the density) becomes 
irrelevant, until it increases enough to draw more than one particle to the 
impurity site, at which point the frozen site again becomes available as
a degree of freedom and ${\cal E}$ is slightly increased. This freezing of
the density distribution at $n=1$ is a strong-interaction effect (it vanishes 
for $U=0$), similar to the Mott insulator in solids, and to Coulomb blockade 
in quantum  dots \cite{qdot,confined}, and to the density plateaus in our 
next example.
  
\begin{figure}
\begin{center}
\includegraphics[height=80mm,width=90mm,angle=0]{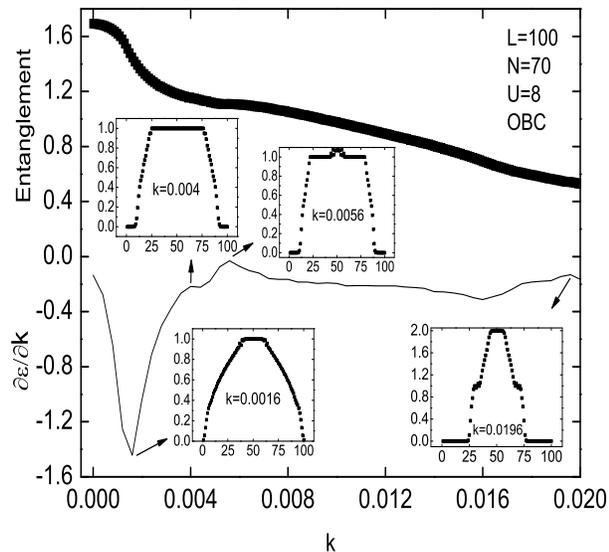}
\end{center}
\caption{\label{fig4} Thick curve: Average per-site entropy 
${\cal E}^{LDA}(n_i,U)/L$ evaluated on BA-LDA densities,
of harmonically trapped fermions on a lattice, as a function of curvature of
the confining potential. Thin curve: numerical derivative of the thick curve. 
Insets: density profiles corresponding to special points on the two curves.}
\end{figure}

Finally, we consider harmonically confined fermions, such as electrons in a 
quantum dot \cite{qdot} or fermionic atoms in an optical trap \cite{opttraps}.
Figure~\ref{fig4} displays the entanglement entropy and its derivative as a 
function of curvature of the trapping potential. Increasing confinement
diminishes entanglement of the confined particles, but not in a uniform 
way. As the curvature becomes larger, the system passes through several states 
with distinct density profiles (see insets), whose nature has been clarified 
elsewhere \cite{trapprb,rigol}. 
Interestingly, each transition corresponds to distinctive features 
in the entanglement entropy, many of which are more clearly visible in its 
derivative: the switching on of the confining potential leads to an initial
drop in entanglement, a spike in the derivative signals the appearance of a
Mott insulating region in the trap center, a smaller peak indicates 
ressurection of a metal-like region in the center, and another broad peak is 
associated with formation of a band-insulating state at the trap center. 
The entanglement entropy can thus be used as a tool for characterization of the
various possible states, even though these are not phases in the thermodynamic 
sense. 

On the other hand, it also becomes clear that the actual value of 
${\cal E}$, and its variation with system parameters, depend in a highly 
nontrivial way on fine details of the system. This dependence is a severe 
obstacle for attempts to quantify and exploit entanglement in real-life 
environments. Note that this issue is conceptually distinct from the more 
commonly discussed problems due to decoherence, which also occurs in 
spatially homogeneous situations.

In summary, we find that for all types of inhomogeneity investigated here
 --- superlattice structures with different modulation patterns, itinerant 
electrons in a metallic wire in the presence of impurities, and phase-separated
states in harmonically confined many-fermion systems ---  the entanglement 
entropy of inhomogeneous systems is strikingly different from that of 
homogeneous systems. The influence of 
spatial inhomogeneity is an essential aspect of entanglement 
in real systems, which cannot be simulated or understood by approaches based 
exclusively on uniform systems. This influence must be understood and 
quantified if entanglement is to be used as a resource for performing quantum 
information processing in actual devices. The entropy LDA and the nested LDA 
concept, proposed here, may be useful tools for this investigation. 

This work was supported by FAPESP, CNPq and CAPES.


\end{document}